%% file: main.tex
\documentclass{Interspeech2024}

\interspeechcameraready

\title{ASoBO: Attentive Beamformer Selection for Distant Speaker Diarization in Meetings} 

\name[affiliation={1,2}]{Théo}{Mariotte}
\name[affiliation={2}]{Anthony}{Larcher}
\name[affiliation={1}]{Silvio}{Montrésor}
\name[affiliation={1}]{Jean-Hugh}{Thomas}

\address{
  $^{1}$LIUM, Institut Claude Chappe, Le Mans Université, France\\
  $^{2}$LAUM IA-GS UMR CNRS 6613, Le Mans Université, France}

\email{theo.mariotte@univ-lemans.fr}


\usepackage{comment}
\usepackage{adjustbox}
\usepackage{multirow}
\usepackage{cite}
\begin{document}

\maketitle
 
\begin{abstract}
Speaker Diarization (SD) aims at grouping speech segments that belong to the same speaker. This task is required in many speech-processing applications, such as rich meeting transcription. In this context, distant microphone arrays usually capture the audio signal. Beamforming, i.e., spatial filtering, is a common practice to process multi-microphone audio data. However, it often requires an explicit localization of the active source to steer the filter. This paper proposes a self-attention-based algorithm to select the output of a bank of fixed spatial filters. This method serves as a feature extractor for joint Voice Activity (VAD) and Overlapped Speech Detection (OSD). The speaker diarization is then inferred from the detected segments. The approach shows convincing distant VAD, OSD, and SD performance, e.g. 14.5\% DER on the AISHELL-4 dataset. The analysis of the self-attention weights demonstrates their explainability, as they correlate with the speaker's angular locations.
\end{abstract}


\input{01_intro}
\input{02_method}
\input{03_protocol}
\input{04_results}
\input{05_analysis}

\input{06_ccl.tex}

\section{Ackowledgments}
This project has received funding from the European Union’s Horizon 2020 research and innovation program under the Marie Skłodowska-Curie grant agreement No 101007666. 
This work was performed using HPC resources from GENCI–IDRIS (Grant 2022-AD011012565).

\bibliographystyle{IEEEtran}
\bibliography{mybib}

\end{document}

%% file: 01_intro.tex
\section{Introduction}

Speaker diarization (SD) is an automatic speech processing task that answers the question \textit{Who spoke and when?} in an audio stream.
It is of major interest for rich meeting transcriptions where the speaker activity is required \cite{yu2022m2met}.
Two categories of systems appear in the literature \cite{park2022review}: end-to-end neural diarization (EEND) \cite{fujita2019end,horiguchi2022encoder} and pipeline systems \cite{bredin_pyannoteaudio_2020,landini2022bayesian}.
The former infers speaker activities from the raw audio signal and usually requires large synthetic datasets to be trained.
The latter comprises sub-blocks that (1) detect speaker-homogeneous segments, and (2) cluster these segments to group them by speakers.
The segmentation step can be divided into two tasks: Voice Activity Detection (VAD) to detect speech segments \cite{gelly2017optimization,lavechin2019end}, and Overlapped Speech Detection (OSD) to identify segments where several speakers are simultaneously active \cite{bullock_overlap-aware_2020,lebourdais22_interspeech}.
Speaker change detection (SCD) \cite{yin2017speaker} can also be performed to detect boundaries between speakers in speech segments.
However, this task is out of the scope of this paper.

SD in meetings is a challenging task due to spontaneous speech, an unknown number of speakers, and difficult acoustic conditions \cite{wolfel2009distant,anguera2007acoustic}.
This scenario remains challenging, as shown by the recently organized challenges \cite{yu2022m2met,vinnikov2024notsofar}.
A common approach is to record meetings with a multi-microphone device \cite{yu2022m2met,carletta2006ami,fu21b_interspeech} such as uniform circular arrays (UCA)\cite{benesty2008microphone,benesty_design_2015}.
Microphone arrays have been widely studied in the literature \cite{benesty2008microphone,benesty_design_2015}.
Specifically, beamforming extracts a signal steered in a given direction, e.g. by weighting and combining channels in the Fourier domain \cite{benesty_design_2015}.
Both signal- \cite{anguera2007acoustic,benesty_design_2015,benesty2008microphone,gannot2001signal} and neural-based \cite{ochiai2017unified,heymann2017beamnet,cornell2022learning} approaches have been investigated.
While neural beamformers increase the number of trainable parameters, signal-based approaches often require estimating the source direction of arrival (DoA).

In this work, we introduce the Attentive Selection of Beamfomer Outputs (ASoBO) as a multi-microphone front-end for VAD+OSD. 
A set of signal-based beamformers is steered in fixed, separated angular directions.
The outputs of the beamformers are weighted and combined with a Self-Attention Channel Combinator (SACC) \cite{gong_self-attention_2021}.
This results in a single-channel enhanced representation of the multi-microphone input signal.
ASoBO prevents the DoA estimation step while limiting the number of trainable parameters.
A first SD is obtained by first applying the VBx system \cite{landini2022bayesian} to the VAD output.
The final SD is inferred by detecting and assigning OSD segments \cite{otterson2007efficient,landini2021analysis}.
The proposed ASoBO shows convincing SD performance on two multi-microphone datasets recorded in the meeting scenario.
Furthermore, we show that the speaker's angular direction can be inferred in an unsupervised way from the self-attention weights.
The code is available at \url{https://git-lium.univ-lemans.fr/speaker/sidiar/}.


%% file: 02_method.tex
\vspace{-5pt}
\section{Segmentation for speaker diarization}
\begin{figure}[t]
    \centering
    \includegraphics[width=\linewidth]{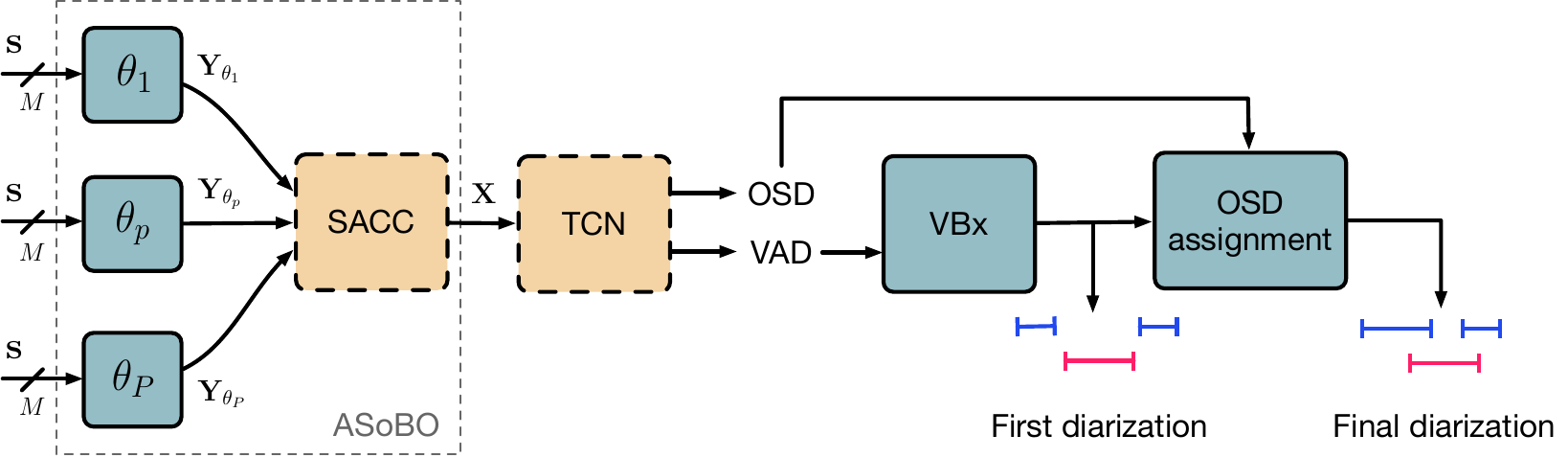}
    \caption{High-level architecture diagram of the proposed ASoBO and the speaker diarization pipeline. The $\theta_p$ blocks represent the fixed spatial filters. Only dash line blocks are trained.}
    \label{fig:archi}
    \vspace{-15pt}
\end{figure}

The proposed speaker diarization (SD) system is pipeline-based. 
This section describes the VAD+OSD formulation and its use for SD.
The overall architecture is presented in figure \ref{fig:archi}.

\vspace{-5pt}
\subsection{Feature extraction}

Let $\mathbf{X}=[\mathbf{X}_1,\dots \mathbf{X}_t, \dots \mathbf{X}_{T}]\in \mathbb{R}^{F\times T}$ be a sequence of feature vectors where $F$ is the number of features, $T$ the number of time frames and $t$ the time frame index. 
This sequence is extracted from the raw audio signal $\mathbf{s}\in\mathbb{R}^{M\times L}$,  with $M$ being the number of microphones and $L$ being the number of samples.
Feature extraction can be defined as a function $g:\mathbb{R}^{M\times L}\rightarrow\mathbb{R}^{F\times T}$, which maps the raw input signal to the sequence of feature vectors.
In this paper, we propose a new design for the $g$ function that uses a set of beamformers, i.e. spatial filter banks, followed by a self-attention model to combine the filter bank output channels.
The method is presented in Section \ref{sect2_ch_feat}.

\subsection{Frame classification}

The sequence of feature vectors serves as input for a VAD+OSD system.
Let $\mathbf{y}=[y_1,\dots y_t,\dots y_{T}]\in \mathbb{R}^{T}$ be a sequence of reference binary labels aligned with the sequence $\mathbf{X}$. 
VAD+OSD is solved by optimizing the parameters $\hat{\boldsymbol{\theta}}$ of the model $f : \mathbf{X}, \boldsymbol{\theta} \rightarrow \hat{\mathbf{y}}$ which maps the feature sequence to a sequence of predicted labels $\hat{\mathbf{y}}=[\hat{\mathbf{y}}_1,\dots\hat{\mathbf{y}}_t,\dots\hat{\mathbf{y}}_{T}]\in \mathbb{R}^{C\times T}$, with $C$ being the number of classes.
An element of $\mathbf{\hat{y}}_t$ contains the pseudo-probability for the frame $\mathbf{X}_t$ to belong to each class. 
The system is trained to predict $C=3$ classes.
The first class corresponds to the non-speech scenario with $N_{spk}=0$ active speaker.
The second and third classes correspond to $N_{spk}=1$ and, $N_{spk}\geq 2$ respectively.
Therefore, VAD can be solved by combining the two last outputs, i.e. $N_{spk}>0$.
OSD is inferred from the $N_{spk}\geq 2$ output.

\subsection{Integrating segmentation for speaker diarization}

VAD and OSD predictions are used to solve speaker diarization (SD) as illustrated in figure \ref{fig:archi}.
The speech segments detected with VAD (all speech segments, overlap included) are used to extract speaker embeddings.
The embeddings are then clustered, and the segmentation is refined using the VBx approach \cite{landini2022bayesian}.
Once the first SD, OSD can be used to assign a second speaker to overlapping speech regions by an additional post-processing step.
In this work, we use the approach from \cite{otterson2007efficient} that assigns the closest speaker in time to the overlapping segment.
This approach has shown on-par performance as more complex approaches such as VB-based methods \cite{landini2021analysis}.

\section{Self-Attentive beamformer selection}
\label{sect2_ch_feat}

This section describes the proposed feature extraction algorithm. 
An abstract view of the method is depicted in figure \ref{fig:archi}.

\subsection{Super-directive beamforming}

Super-directive beamforming is a commonly used algorithm for spatial filtering \cite{wolfel2009distant}. 
The narrowband weights of such a filter can be expressed as follows:

\begin{equation}
    \mathbf{w}_p^H(f)=\frac{\mathbf{v}_p^H(f)\boldsymbol{\Sigma}_N^{-1}(f)}{\mathbf{v}_p^H(f)\boldsymbol{\Sigma}_N^{-1}(f)\mathbf{v}_p(f)},
    \label{eq:chp3:sdbf_weights}
\end{equation}
where $f$ is the frequency, $\mathbf{v}_p(f)\in\mathbb{C}^{M\times 1}$ a steering vector and $\boldsymbol{\Sigma}_N(f) \in \mathbb{R}^{M\times M}$ the noise covariance matrix.
In this work, we use the standard isotropic noise assumption usually considered in super-directive beamforming \cite{wolfel2009distant}.
When considering a UCA, the $m$-th element of the steering vector $v_{p,m}$, oriented towards the $\theta_p$ angular direction, can be expressed as \cite{benesty_design_2015} :
$v_{p,m}(f) = \exp{\left( j2\pi frc^{-1} \cos (\theta_p-\psi_m)\right)}, $ 
with $m$ being the index of the microphone with angle $\psi_m$, $c$ the speed of sound and $r$ the radius of the UCA.

Let $\mathbf{S}\in\mathbb{C}^{M\times F\times T}$ be the short-time Fourier Transform (STFT) of the multi-microphone input signal $\mathbf{s}$.
The output of the $p$-th filter -- steered in the $\theta_p$ direction -- at frequency $f$, is obtained following
\begin{equation}
    \mathbf{Y}_p(t,f) = \mathbf{w}_{p}^{H}(f)\mathbf{S}(t,f).
    \label{eq:chp3:bf_output}
\end{equation}
The output $\mathbf{Y}_p\in\mathbb{C}^{T\times F}$ is a single-channel signal steered towards the $\theta_p$ direction.
One can build a set of spatial filters $\mathcal{W}=\{\mathbf{W}_p\}_{p=1}^P$ steered in $P$ unique angular directions, where $\mathbf{W}_p = [\mathbf{w}_{p}(f_i)], i=1,\dots,F$ is the broadband filter coefficients. 
Filtering $\mathbf{S}$ by all the filters from $\mathcal{W}$ results in a new multichannel signal $\mathbf{Y}\in\mathbb{C}^{T\times P\times F}$, where the $p$-th channel corresponds to the beamformed version of $\mathbf{S}$ in the $\theta_p$ direction.


\subsection{Beamformer selection}

Once the input signal is filtered by the spatial filter bank $\mathcal{W}$, a second step selects the optimal directions at the frame level.
This filter selection is performed using the Self-Attention Channel Combinator (SACC) module, which is efficient for audio channel selection \cite{gong_self-attention_2021,mariotte2024channel}.
Contrary to \cite{mariotte2024channel}, the SACC is applied to beamforming outputs instead of the multi-microphone signal directly.
We first only keep the magnitude of the beamforming output $\mathbf{Y}$.
Then, $|\mathbf{Y}|^2$ is projected by three linear layers to the query and key $\mathbf{Q},\mathbf{K}\in\mathbb{R}^{T\times P\times D}$, and the value $\mathbf{V}\in\mathbb{R}^{T\times P\times 1}$, with $D$ being the output dimension of the linear transformation.
The attention weights $\mathbf{w}_{SA}\in\mathbb{R}^{T\times P}$ are computed as follows:

\begin{equation}
    \mathbf{w}_{SA}=\mathrm{softmax}\left(\frac{\mathbf{Q}\mathbf{K}^T}{\sqrt{D}}\right)\mathbf{V}, 
    \label{eq:self_att_weights}
\end{equation}
with $\cdot^T$ the transpose operator applied to each frame of $\mathbf{K}$.
The spectrogram resulting from the attentive selection $\bar{\mathbf{Y}}$ is calculated by first weighting $\mathbf{Y}$ with $\mathbf{w}_{SA}$.
A sum is then applied on the channel dimension $P$.
For a given frame $t$, this operation can be expressed as:

\begin{equation}
    \bar{\mathbf{Y}}_t = \sum_{p=1}^P\mathrm{softmax}\left(\mathbf{w}_{SA,t}\right) \odot \mathbf{Y}_t,
    \label{eq:comb_channels}
\end{equation}
with $\odot$ being the element-wise product on the channel dimension, i.e. one weight at $t$ is applied to all the frequencies of $\mathbf{Y}$.
The $\mathrm{softmax}$ activation function is applied to the channel dimension such that $w_{SA,t,p}\in[0,1]$.
The final feature sequence $\mathbf{X}$ is obtained by converting $\bar{\mathbf{Y}}$ to the Mel scale with 64 triangular filters \cite{gong_self-attention_2021}.

%% file: 03_protocol.tex
\section{Experimental protocol}
\label{sect3_protocol}

\subsection{Datasets}

The experiments are conducted on two datasets featuring distant multi-microphone speech with known array geometry: AMI \cite{carletta2006ami} and AISHELL-4 \cite{fu21b_interspeech}.
The AMI corpus is about 100h of meetings in English with up to 5 participants recorded using different devices.
In this work, we use the audio recorded by the 8-microphone, 10cm-radius UCA placed in the center of the table during the sessions.
The data partition follows the protocol proposed in \cite{landini2022bayesian} since it guarantees no speaker overlap between the subsets.
The AISHELL-4 dataset provides 120 hours of conference recordings with 4 to 8 participants.
Audio is recorded with an 8-microphone, 5cm-radius UCA usually placed in the center of the table.
Meetings are in Mandarin and were recorded in various acoustic environments.
Both datasets are sampled at 16kHz.

\subsection{Implementation details}

The ASoBO inputs complex STFT extracted on 25ms segments with 10ms shift.
$P$ spatial filters, steered in uniformly-spaced angular sectors between 0 and 2$\pi$ are applied to the input STFT.
We empirically found that $P=4$ and $P=8$ offer the best performance on AMI and AISHELL-4, respectively.
The $P$-channel output signal is then processed by the SACC algorithm with a hidden size $D=256$.
The modeling of the ASoBO feature sequence is performed with the same TCN-based architecture as \cite{Cornell2022}.
It is composed of 3 TCN blocks with residual connections.
Each block contains 5 1D convolutional layers with exponentially increasing dilation.

The speaker diarization is inferred with the VBx implementation proposed in \cite{landini2022bayesian}.
This system uses a ResNet101 x-vector extractor followed by a VB-HMM clustering algorithm.
We used the default AMI diarization setup from the available code\footnote{\url{https://github.com/BUTSpeechFIT/VBx}}.
The VAD segments, predicted by our systems, are used as an initial segmentation.
X-vector clustering is initialized with Hierarchical Agglomerative Clustering (HAC) before performing VB clustering.
Overlapped speech segments are assigned as a post-processing step using a heuristic approach \cite{otterson2007efficient}.

\subsection{Baselines}

The ASoBO approach is compared to two baseline systems.
As a lower-bound system, we consider the single-distant microphone (SDM) scenario.
The segmentation is performed on the first microphone of the array.
20 Mel Frequency Cepstral Coefficients (MFCC) and its deltas are extracted from the audio signal on 25ms windows with 10ms shift.
These features are fed directly to the segmentation system to obtain VAD and OSD predictions.
As an upper-bound baseline, we consider the original implementation of the SACC architecture, which has shown strong OSD performance in the multichannel distant scenario \cite{mariotte2024channel}.
The SACC is applied to the magnitude of the STFT calculated on 25ms windows with 10ms shift.
The self-attention hidden dimension is set to $D=256$.
The speaker diarization performance is compared to \cite{raj2022gpu} on both AMI and AISHELL-4 datasets.
They report the performance obtained with both VBx and spectral clustering approaches.

\subsection{Training and evaluation}

The segmentation systems are trained on 200 epochs with batches of 64 segments.
The training segment duration is fixed to 2s.
The segmentation is solved as a multiclass classification task and is optimized in a supervised way using the cross-entropy loss.
The weights of the models are optimized with the ADAM optimizer with the learning rate set to 0.001.

The evaluation is conducted on the evaluation set of both datasets.
VAD and OSD predictions are inferred from a 2s sliding window with a 0.5s shift.
VAD is evaluated regarding False Alarm (FA) and Missed Detection (MD) rates.
The sum of both metrics, the Segmentation Error Rate (SER) is also reported \cite{lavechin2019end}.
OSD is evaluated in terms of Precision (P), Recall (R), and F1-score (F1).
The speaker diarization is evaluated using the Diarization Error Rate (DER) \cite{park2022review}.
We report the scores with ($\delta=0.25)$, and without ($\delta=0$) forgiveness collar.
Unless otherwise specified, values highlighted in bold indicate the best systems and the statistically equivalent ones ($p<0.001$).
We use the Wilcoxon signed-rank non-parametric test on the file-level scores \cite{demvsar2006statistical}.

%% file: 04_results.tex
\section{Experimental study}
\label{sect4_seg_perf}
This section presents the experimental results on both segmentation (VAD+OSD) and speaker diarization.

\subsection{Segmentation performance}

Table \ref{tab:seg_perf} presents the VAD and OSD performance on both AMI and AISHELL-4 datasets.
On the AMI corpus, the SDM system reaches 6.57\% SER on VAD and 65.4\% F1-score on OSD.
SACC outperforms the SDM model with 5.59\% VAD SER and 68.4\% OSD F1-score. 
The proposed ASoBO system shows mitigated performance with 6.53\% SER on VAD but improves OSD compared to SDM with 67.2\% F1-score.
The VAD degradation can be explained by the high false alarm rate (4.16\%).
On the AISHELL-4 dataset, ASoBO reaches the best VAD performance with 4.39\% SER compared to the SACC (4.57\%) and the SDM (5.17\%).
The OSD scores on this dataset are low for each model.
This can be explained by the low quality of the annotations on overlapping speech.
The SDM shows a 31.3\% F1-score and is largely outperformed by SACC (41.1\%) and ASoBO (40.5\%).
In summary, the ASoBO system improves VAD+OSD concerning the SDM scenario.
The segmentation performance on the AMI corpus is still mitigated compared to the original SACC.
However, this approach also improves the segmentation on the AISHELL-4 dataset.


\begin{table}[t]
    \centering
    \caption{VAD+OSD performance on the AISHELL-4 and AMI evaluation sets. \# Param. represents the number of trainable parameters in millions.}
    \begin{adjustbox}{max width=\linewidth}
    \begin{tabular}{lccccccc}
    \toprule
    & & \multicolumn{3}{c}{VAD} & \multicolumn{3}{c}{OSD} \\
    \cmidrule(lr){3-5}
    \cmidrule(lr){6-8}
    AMI & \#Param.  & {FA} & {Miss} & {SER} & {P} & {R} & {F1} \\
    \midrule
    SDM & 0.26M & 4.33 & 2.24 & 6.57 & 73.8 & 68.8 & 65.4 \\
    SACC & 0.40M & 2.91 & 3.61 &\textbf{5.59} & 78.1 & 60.8 & \textbf{68.4} \\
    ASoBO & 0.36M & 4.16 & 2.15 & 6.53 &  70.8 & 69.3 & 67.2\\
    \midrule
    \multicolumn{2}{l}{AISHELL-4}  & {FA} & {Miss} & {SER} & {P} & {R} & {F1} \\
    \midrule
    SDM & 0.26M & 3.69 & 1.48 & 5.17 & 20.4 & 67.1 & 31.3 \\
    SACC & 0.40M & 3.35 & 1.21 & 4.57 & 28.4 & 74.4 & \textbf{41.1}\\
    ASoBO & 0.36M & 2.29 & 2.10 & \textbf{4.39} & 28.7 & 69.0 & \textbf{40.5} \\

    \bottomrule
    \end{tabular}
    \end{adjustbox}
    \label{tab:seg_perf}
    \vspace{-10pt}
\end{table}

\subsection{Speaker diarization performance}

Table \ref{tab:diar_perf} shows the speaker diarization performance on both AMI and AISHELL-4 evaluation sets.
The VBx-based system from \cite{raj2022gpu} reaches 25.1\% and 18.0\% DER on AMI and AISHELL-4 respectively.
The spectral clustering-based model offers 23.7\% and 16.1\% on these datasets.

On the AMI corpus, the SACC offers the best speaker diarization performance with 23.1\%.
Note that the overlap assignment improves the diarization performance by a relative +11.5\%.
ASoBO offers close performance with 24.1\% when OSD segments are assigned.
This system is limited by the segmentation performance. 
It reaches 16.7\% DER when a forgiveness collar is applied, which is 0.4\% far from SACC.
Both SACC and ASoBO improve or offer similar performance as the baseline.
The SDM system reaches 25.0\% DER and is largely outperformed by multichannel systems.

The trend is different on the AISHELL-4 dataset.
First, the OSD segment assignment degrades the performance.
This was expected based on the low OSD performance on this dataset and was also observed in \cite{raj2022gpu}.
On this dataset, both SDM and SACC offer close performance, with 16.7\% and 16.4\% DER.
ASoBO reaches the best DER with 14.5\%.
It improves SACC by a relative +11.5\%.

In summary, ASoBO is a good candidate for distant speaker diarization with pipeline systems.
While the improvement in the AMI data is mitigated, the performance on the AISHELL-4 dataset is noticeable and very encouraging for such a system.
Beyond performance, self-attentive selection of the beamformer makes ASoBO an explainable system as shown in section \ref{sect5:local}.

\begin{table}[t]
    \centering
    \caption{Diarization Error Rate (DER) with each segmentation system on AMI and AISHELL-4 evaluation sets.}
    \begin{adjustbox}{max width=\linewidth}
    \begin{tabular}{lcccc}
    \toprule
    & \multicolumn{2}{c}{AMI} & \multicolumn{2}{c}{AISHELL-4} \\
    \cmidrule(lr){2-3}
    \cmidrule(lr){4-5}
        & $\delta=0.25$ & $\delta=0$ & $\delta=0.25$ & $\delta=0$\\
    \midrule
    VBx \cite{raj2022gpu} & - & 25.1 & - &  18.0\\
    Spectral \cite{raj2022gpu} & - & 23.7 & - &  16.1\\
    \midrule
    SDM & 19.4 & 27.5 & 11.3 & 16.7\\
    $\hookrightarrow$ w/ OSD & 17.7 & 25.0 & 24.9 & 29.3\\
    SACC & 18.2 & 26.1 & 11.0 & 16.4\\
    $\hookrightarrow$ w/ OSD & \textbf{16.3} & \textbf{23.1} & 19.3 & 23.9\\
    ASoBO & 18.6 & 26.9 & \textbf{9.2} & \textbf{14.5}\\
    $\hookrightarrow$ w/ OSD & \textbf{16.7} & 24.1 & 16.6 & 20.9\\
    \bottomrule
    \end{tabular}
    \end{adjustbox}
    \label{tab:diar_perf}
    \vspace{-10pt}
\end{table}

%% file: 05_analysis.tex
\section{Weight explanation}
\label{sect5:local}
The self-attention module is trained to select the appropriate filters for the segmentation task.
The intuition is that the self-attention model selects the filter steered toward the active speakers.
This section verifies this hypothesis by analyzing the combination weights on simulated data.

\subsection{Data simulation}

The AMI and AISHELL-4 datasets do not come with speaker position annotations.
The analysis of self-attention weights is performed on a simulated dataset.
The simulations are conducted using the LibriMix dataset \cite{cosentino2020librimix} featuring single-channel speech mixtures. 
The original mixtures are spatialized with simulated Room Impulse Responses (RIRs).
These are generated using the GpuRIR toolkit \cite{diaz-guerra_gpurir_2021}.
The simulation setup is similar to the AMI corpus, with an 8-microphone UCA with a 10cm radius.
We simulate the 2- and 3-speaker mixtures, with a reverberation time $T_{60}$ of 0.6 seconds.
Two evaluation scenarios are considered: \textit{easy} where the sources are always aligned with a beamformer, \textit{i.e.} one of the $P$ directions, and \textit{hard} where the source position is randomly sampled around the $p$-th selected filter direction with $\pm5^\circ$.

\begin{figure}[b]
    \centering
    \includegraphics[width=0.9\linewidth]{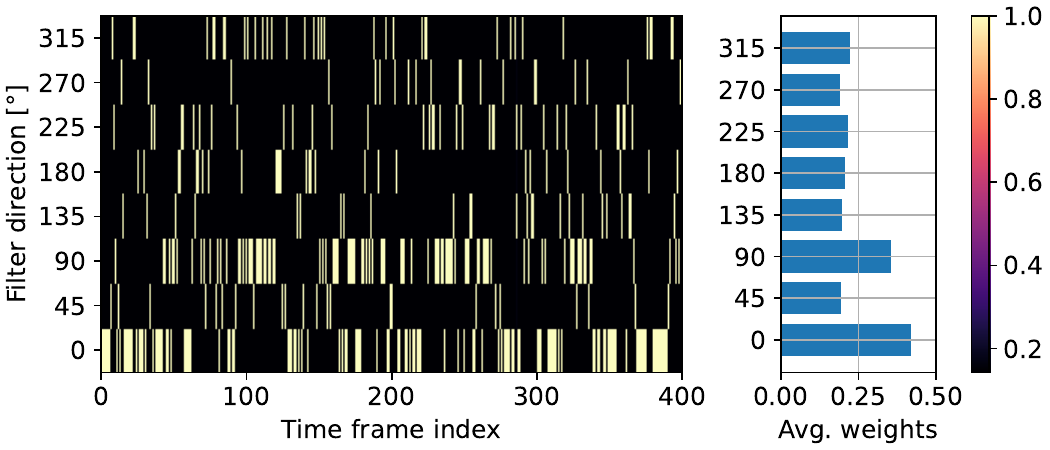}
    \caption{(left) Combination weights for a spatialized utterance of Libri2Mix with speakers located at 0$^\circ$ and $90^\circ$. (right) Time-averaged weights from the same utterance.}
    \label{fig:selection_weights}
    \vspace{-15pt}
\end{figure}

\subsection{Speaker localization from combination weights}

Identifying the steering directions of the system can be seen as a speaker localization task.
Let $\mathbf{w}_{SA}\in\mathbb{R}^{T\times P}$ be a set of ASoBO combination weights predicted from a given audio segment, calculated with equation \eqref{eq:self_att_weights}.
The activations in $\mathbf{w}_{SA}$ indicate the selected angular directions as a function of time.
The average steering direction is calculated by first averaging $\mathbf{w}_{SA}$ across time:
$\bar{\mathbf{w}} = \frac{1}{T}\sum_{t=1}^T\mathrm{softmax}\left(\mathbf{w}_{SA}\right)_t,$
where the $\mathrm{softmax}$ activation is applied on the channel dimension as in equation \eqref{eq:comb_channels}.
Hard labels are obtained by applying a threshold $\tau\in[0,1]$ to each element of $\bar{\mathbf{w}}$.
The angular directions $\hat{\theta}_p$ can be compared to the ground truth directions $\theta_p$ using the precision (P), recall(R) and F1-score (F1) metrics.


Figure \ref{fig:selection_weights} illustrates the combination weights obtained for two active sources at 0$^\circ$ and 90$^\circ$ respectively with the ASoBO system with $P=8$ filters.
The weight map (left) shows that the two directions of the active speaker are more often activated than the others. 
The averaged weights (right) confirm this behavior, with two peaks in the speakers' directions.


\begin{table}[t]
    \centering
    \caption{Speaker localization performance of ASoBO on spatialized Libri2Mix and Libri3mix development sets. $L$ represents the number of simultaneously active sources.}
    \begin{adjustbox}{max width=\linewidth}
    \begin{tabular}{cccccccc}
        \toprule
        & \multicolumn{3}{c}{$L=2$} & \multicolumn{3}{c}{$L=3$}\\
        \cmidrule(lr){2-4}
        \cmidrule(lr){5-7}
         Scenario & P & R & F1 & P & R & F1 \\
         \midrule
         \textit{Random} & 49.9 & 49.9 & 49.9 & 49,7 & 49,6 & 48,7\\
         \textit{Easy}  & 84.0 & 83.2 & 83.6 & 82.5 & 76.1 & 75.9\\
         \textit{Hard} & 70.4 & 73.0 & 71.7 & 66.9 & 66.5 & 66.6\\
         \bottomrule
    \end{tabular}
    \end{adjustbox}
    \vspace{-10pt}
    \label{tab:localization}
\end{table}

\subsection{Localization performance}

The spatial filter selection is evaluated as a speaker localization task.
Each evaluation is conducted on a spatialized version of the LibriMix development set.
Localization performance in the \textit{easy} and \textit{hard} scenarios, for 2- and 3-source mixtures, are presented in Table \ref{tab:localization}.
This analysis is performed on the $P=8$-filter ASoBO system trained on AISHELL-4.
The \textit{random} row corresponds to the random selection of the filters.
For $L=2$ sources, the system tends to select the closest direction to the speaker.
This is shown by the 83.6\% and 71.7\% F1-score in the \textit{easy} and \textit{hard} scenarios, respectively.
Note that the localization score is strongly degraded when the source is not aligned with the speaker.
For $L=3$, the system still localizes the sources accurately.
The \textit{easy} scenario shows a 75.9\% F1-score.
This score degrades--but remains higher than random selection--in the \textit{hard} case, with 66.6\%.

This study shows how the self-attention module can select the filter direction associated with the active speaker.
The degradation in the \textit{hard} scenario can explain the diarization performance limitation of our system on the AMI dataset.
If the active speakers are not aligned with the filters, the self-attention might extract the features from the wrong spatial filter.

%% file: 06_ccl.tex
\section{Conclusions}
\label{sect_ccl}
In this paper, we propose a multi-microphone segmentation algorithm for distant speaker diarization.
This method consists of a set of beamformers steered in fixed directions whose outputs are selected with a self-attention module.
The output of this system serves as a feature sequence for a joint Voice Activity (VAD) and Overlapped Speech Detection (OSD) system.
This segmentation is used for speaker diarization with the VBx system.
Experiments on the AMI and AISHELL-4 datasets have shown that the proposed approach improves the speaker diarization under distant conditions, with mitigated gain on AMI but significant improvement on AISHELL-4.
The analysis of the weights of the self-attention model shows that we can perform pseudo-localization of the active speakers.
This demonstrates how explainable such a system can be.
In future work, we plan to evaluate the impact of mismatched array setup on the speaker diarization performance and to estimate the Direction of Arrival at inference time.